%% file: canari.tex
\documentclass[]{spie}  
\usepackage[]{graphicx}

\title{Quantum fluctuations and glassy behavior of electrons near metal-insulator transitions}

\author{V. Dobrosavljevi\'{c}
\skiplinehalf Department of Physics and National High Magnetic
Field Laboratory, Florida State University, Tallahassee, FL 32306,
USA}

\authorinfo{Further author information: E-mail: vlad@magnet.fsu.edu}

  \include{mydefs}

\begin{document}
  \maketitle

\begin{abstract}
Glassy behavior is a generic feature of electrons close to
disorder-driven metal-insulator transitions. Deep in the
insulating phase, electrons are tightly bound to impurities, and
thus classical models for electron glasses have long been used. As
the metallic phase is approached, quantum fluctuations become more
important, as they control the electronic mobility. In this paper
we review recent work that used extended dynamical mean-field
approaches to discuss the influence of such quantum fluctuations
on the glassy behavior of electrons, and examine how the stability
of the glassy phase is affected by the Anderson and the Mott
mechanisms of localization.
\end{abstract}



\section{Glassy behavior as a precursor to the metal-insulator transition}
\label{sect:intro}  

Understanding the metal-insulator transition (MIT) poses one of
the most basic questions of condensed matter physics.  It has been
been a topic of much controversy and debate starting from early
ideas of Mott \cite{re:Mott90}, and Anderson \cite{re:Anderson58},
but the problem remains far from being resolved. Quite generally,
when a system is neither a good metal nor a good insulator, both
the localized and the itinerant aspects of the problem are
important. In this intermediate regime, several competing
processes can be simultaneously present.
   \begin{figure}[h]
   \begin{center}
   \begin{tabular}{c}
   \includegraphics[width=4in]{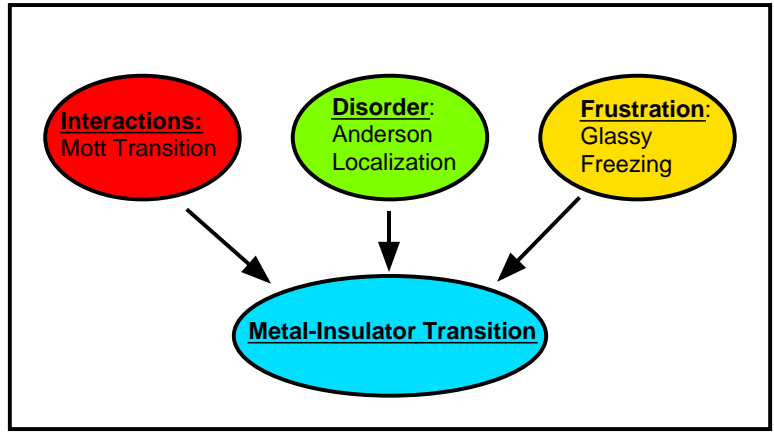}
   \end{tabular}
   \end{center}
   \caption{Three basic routes to localization}
   \end{figure}
As a result, the system cannot ``decide'' whether to be a metal or
an insulator until a very low temperature $T^*$ is reached, below
which a more conventional description applies.  This situation is
typical of systems close to a quantum critical point
\cite{re:Sondhi97}, which describes a zero temperature second
order phase transition between two distinct states of matter.
Understanding the nature of low energy excitations in the
intermediate regime between a metal and an insulator is of crucial
importance for the progress in material science.

The primary reason for theoretical difficulties is related to the
fact that both the Mott and the Anderson transition find
themselves in regimes where traditional, perturbative approaches
\cite{re:Lee85} cannot be straightforwardly applied. To make the
problem even more difficult, simple estimates \cite{re:Mott90} are
sufficient to appreciate that in many situations the effects of
interactions and disorder are of comparable magnitude and thus
both should be simultaneously considered.  So far, very few
approaches have attempted to simultaneously incorporate these two
basic routes to localization.

Another aspect of disordered interacting electrons poses a
fundamental problem. Very generally, Coulomb repulsion favors a
uniform electronic density, while disorder favors local density
fluctuations. When these two effects are comparable in magnitude,
one can expect many different low energy electronic
configurations, i.e. the emergence of many {\em metastable
states}. Similarly as in other ``frustrated'' systems with
disorder, such as spin glasses, these processes can be expected to
lead to {\em glassy} behavior of the electrons, and the associated
anomalously slow relaxational dynamics. Indeed, both theoretical
\cite{eglass1,eglass2} and experimental
\cite{films21,films22,films23,films24,films25} work has found
evidence of such behavior deep on the insulating side of the
transition. However, at present very little is known as to the
precise role of such processes in the critical region.
Nevertheless, it is plausible that the glassy freezing of the
electrons must be important, since the associated slow relaxation
clearly will reduce the mobility of the electrons.  From this
point of view, the glassy freezing of electrons may be considered,
in addition to the Anderson and the Mott mechanism, as a third
fundamental process associated with electron localization.
Interest in understanding the glassy aspects of electron dynamics
has experienced a genuine renaissance in the last few years,
primarily due to experimental advances. Emergence of many
metastable states, slow relaxation and incoherent transport have
been observed in a number of strongly correlated electronic
systems. These included transition metal oxides such as high Tc
materials, manganites, and ruthenates. Similar features have
recently been reported in two-dimensional electron gases, and even
three dimensional doped semiconductors such as Si:P.

\section{Extended DMFT approaches for disordered electrons}

A number of experimental and theoretical investigations have
suggested that the conventional picture of disordered interacting
electrons may be incomplete.  Most remarkably, the characteristic
``critical'' behavior seen in many experiments covers a
surprisingly broad range of temperatures and densities.  This is
more likely to reflect an underlying ``mean-field'' behavior of
disordered interacting electrons than the asymptotic critical
behavior described by an effective long-wavelength theory.
\begin{figure}[h]
\begin{center}
\includegraphics[width=4in]{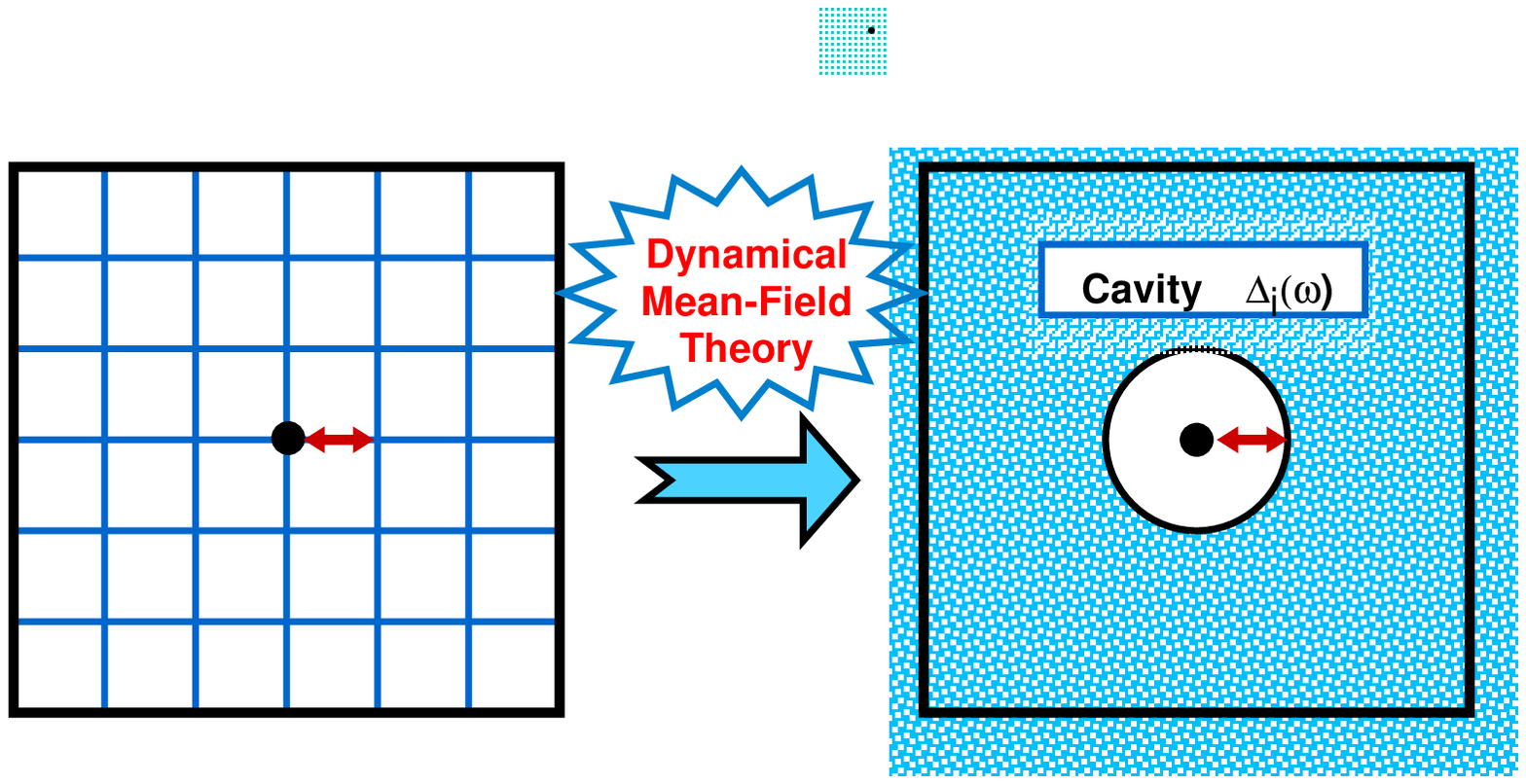}
\caption{In dynamical mean-field theory, the environment of a
given site is represented by an effective medium, represented by
its ``cavity spectral function'' $\Delta_i (\omega )$. In a {\em
disordered} system, $\Delta_i (\omega )$ for different sites can
be very different, reflecting Anderson localization effects.}
\end{center}
\end{figure}
Thus a simple mean-field description is needed to provide the
equivalent of a Van der Waals equation of state, for disordered
interacting electrons.  Such a theory has long been elusive,
primarily due to a lack of a simple order-parameter formulation
for this problem.  Very recently, an alternative approach to the
problem of disordered interacting electrons has been formulated,
based on dynamical mean-field  theory (DMFT) methods
\cite{re:Georges96}. This formulation is largely complementary to
the scaling approach, and has already resulting in several
striking predictions.

 The DMFT approach focuses on a single lattice site, but replaces
\cite{re:Georges96} its environment by a self-consistently
determined ``effective medium'', as shown in Fig. 2. For itinerant
electrons, the environment cannot be represented by a static
external field, but instead must contain the information about the
dynamics of an electron moving in or out of the given site. Such a
description can be made precise by formally integrating out
\cite{re:Georges96} all the degrees of freedom on other lattice
sites. In presence of electron-electron interactions, the
resulting local effective action has an arbitrarily complicated
form. Within DMFT, the situation simplifies, and all the
information about the environment is contained in the local single
particle spectral function $\Delta_i (\omega )$.  The calculation
then reduces to solving an appropriate quantum impurity problem
supplemented by an additional self-consistency condition that
determines this ``cavity function'' $\Delta_i (\omega )$.

The precise form of the DMFT equations depends on the particular
model of interacting electrons and/or the form of disorder, but
most applications \cite{re:Georges96} to this date have focused on
Hubbard and Anderson lattice models.  The approach has been very
successful in examining the vicinity of the Mott transition in
clean systems in which it has met spectacular successes in
elucidating various properties of several transition metal oxides
\cite{re:Dobrosavljevic98}, heavy fermion systems, and Kondo
insulators \cite{re:Rozenberg96}.

When appropriately generalized to disordered
systems\cite{re:Dobrosavljevic98}, these methods are able to
incorporate all the three basic mechanisms of electron
localization. In particular, the DMFT approach is able to present
a consistent picture for the glassy behavior of electrons, and
discuss its emergence in the vicinity of metal-insulator
transitions. In this paper we review recent results obtained in
this framework, and discuss their relevance to several
experimental systems.

\section{Simple model of an electron glass}

The interplay of the electron-electron interactions and disorder
is particularly evident deep on the insulating side of the
metal-insulator transition (MIT). Here, both experimental
\cite{re:Massey96} and theoretical studies \cite{re:Efros75} have
demonstrated that they can lead to the formation of a soft
\textquotedblleft Coulomb gap\textquotedblright, a phenomenon that
is believed to be related to the glassy behavior
\cite{films21,films22,films23,films24,films25,films32,films32} of
the electrons. Such glassy freezing has long been suspected
\cite{re:Belitz95} to be of importance, but very recent work
\cite{newgang,Sudip} has suggested that it may even dominate the
MIT behavior in certain low carrier density systems. The classic
work of Efros and Shklovskii \cite{re:Efros75} has clarified some
basic aspects of this behavior, but a number of key questions have
remain unanswered.

 As a simplest
example\cite{re:pd}  displaying glassy behavior of electrons, we
focus on a simple lattice model of spinless electrons with nearest
neighbor repulsion $V$ in presence of random site energies
$\varepsilon_i$ and inter-site hopping $t$, as given by the
Hamiltonian

\begin{equation}
H=\sum_{<ij>} ( -t + \varepsilon_i \delta_{ij})
c^{\dagger}_{i}c_{j}\nonumber + V\sum_{<ij>}c^{\dagger}_{i}c_{i}
c^{\dagger}_{j}c_{j}.
\end{equation}

This model can be solved\cite{re:pd} in a properly defined limit
of large coordination number $z$ \cite{re:Georges96}, where an
extended dynamical mean-field (DMF) formulation becomes exact. We
concentrate on the situation where the disorder (or more generally
frustration) is large enough to suppress any uniform ordering. We
then rescale both the hopping elements and the interaction
amplitudes as $t_{ij} \rightarrow t_{ij}/\sqrt{z}$; \hspace{6pt}
$V_{ij} \rightarrow V_{ij}/\sqrt{z}$. As we will see shortly, the
required fluctuations then survive even in the
$z\rightarrow\infty$ limit, allowing for the existence of the
glassy phase. Within this model:

\begin{itemize}
\item The universal form  of the Coulomb gap\cite{re:Efros75}
proves to be a direct consequence of glassy freezing.

\item The glass phase is identified through the emergence of an
extensive number of metastable states, which in our formulation is
manifested as a replica symmetry breaking instability
\cite{re:Mezard86}.

\item As a consequence of this ergodicity breaking
\cite{re:Mezard86}, the zero-field cooled compressibility is found
to vanish at T=0, suggesting the absence of screening
\cite{re:Efros75} in disordered insulators.

\item The quantum fluctuations can melt this glass even at $T=0$,
but the relevant energy scale is set by the electronic mobility,
and is therefore a nontrivial function of disorder.
\end{itemize}We should stress that although this model allows to
examine the interplay of glassy ordering and quantum fluctuations
due to itinerant electrons, it is too simple to describe the
effects of Anderson localization. These effects require extensions
to lattices with finite coordination, and and will be discussed in
the next section.

For simplicity, we focus on a Bethe lattice at half filling, and
examine the $z\rightarrow\infty$ limit.  This strategy
automatically introduces the correct order parameters, and after
standard manipulations \cite{re:Dobrosavljevic94} the problem
reduces to a self-consistently defined single site problem, as
defined by an the effective action of the form

\begin{eqnarray}
&&S_{eff} (i) =\sum_{a}\int_o^{\beta}\int_o^{\beta }d\tau d\tau
'\;[ c^{\dagger a}_i (\tau )( \delta (\tau -\tau ')
\partial_{\tau} +\varepsilon_i \nonumber
+ t^2 G (\tau,\tau ') )c^{a}_i (\tau ')\\
&& +\frac{1}{2}V^2\delta n_i^{a}(\tau )\chi (\tau ,\tau ') \delta
n_i^{a} (\tau ')] +\frac{1}{2}V^2\sum_{a\neq b}\int_o^{\beta
}\int_o^{\beta }d\tau d\tau ' \; \delta n_i^{a}(\tau )\; q_{ab}\;
\delta n_i^{b} (\tau ').\label{eq:ch2_seff}
\end{eqnarray}
Here, we have used functional integration over replicated
Grassmann fields \cite{re:Dobrosavljevic94} $c_{i}^{a} (\tau )$
that represent electrons on site $i$ and replica index $a$, and
the random site energies $\varepsilon_i$ are distributed according
to a given probability distribution $P(\varepsilon_i )$.  The
operators $\delta n_{i}^{a}(\tau )=( c^{\dagger a}_{i} (\tau
)c_{i}^{a} (\tau ) - 1/2)$ represent the {\em density
fluctuations} from half filling. The order parameters $G (\tau
-\tau ')$, $\chi(\tau -\tau ')$ and $q_{ab}$ satisfy the following
set of self-consistency conditions

\begin{eqnarray}
&&G (\tau -\tau ')=\int d\varepsilon_i P(\varepsilon_i )
< c^{\dagger a}_{i} (\tau )c_{i}^{a} (\tau ')>_{eff},\\
&&\chi (\tau -\tau ')=\int d\varepsilon_i P(\varepsilon_i )
< \delta n^{\dagger a}_{i} (\tau )\delta n_{i}^{a} (\tau ')>_{eff},\\
&&q_{ab}=\int d\varepsilon_i P(\varepsilon_i ) < \delta n^{\dagger
a}_{i} (\tau )\delta n_{i}^{b} (\tau ')>_{eff}.
\end{eqnarray}

\subsection{Order parameters}

In these equations, the averages are taken with respect to the
effective action of Eq. (\ref{eq:ch2_seff}). Physically, the
``hybridization function'' $t^2 G (\tau -\tau ')$ represents the
single-particle electronic spectrum of the environment, as seen by
an electron on site $i$.  In particular, its imaginary part at
zero frequency can be interpreted \cite{re:Dobrosavljevic97b} as
the inverse lifetime of the local electron, and as such remains
finite as long as the system is metallic.  We recall
\cite{re:Dobrosavljevic94} that for $V=0$ these equations reduce
to the familiar CPA description of disordered electrons, which is
exact for $z=\infty$.  The second quantity $\chi(\tau -\tau ')$
represents an (interaction-induced) {\em mode-coupling} term that
reflects the {\em retarded} response of the density fluctuations
of the environment. Note that very similar objects appear in the
well-known mode-coupling theories of the glass transition in dense
liquids \cite{re:Cummins94}.  Finally the quantity $q_{ab}$
$(a\neq b)$ is nothing but the familiar Edwards-Anderson order
parameter $q_{EA}$. Its nonzero value indicates that the time
averaged electronic density is spatially non-uniform.

\subsection{Equivalent Infinite Range model}

From a technical point of view, a RSB analysis is typically
carried out by focusing on a free energy expressed as a functional
of the order parameters. In our Bethe lattice approach, one
directly obtains the self-consistency conditions form appropriate
recursion relations \cite{re:Dobrosavljevic94}, without invoking a
free energy functional. However, we have found it useful to map
our $z=\infty$ model to another {\em infinite range} model, which
has {\em exactly} the same set of order parameters and
self-consistency conditions, but for which an appropriate free
energy functional can easily be determined. The relevant model is
still given Eq. (1), but this time with {\em random} hopping
elements $t_{ij}$ and {\em random} nearest-neighbor interaction
$V_{ij}$, having zero mean and variance $t^2$, and $V^2$,
respectively. For this model, standard manipulations
\cite{re:Dobrosavljevic94} result in the following free energy
functional

\begin{eqnarray}
F [G , \chi , q_{ab}] =
&-&\frac{1}{2}\sum_{a}\int_o^{\beta}\int_o^{\beta }d\tau d\tau'
\;[ t^2 G^2 (\tau ,\tau ' ) + V^2 \chi^2 (\tau ,\tau ' )]
-\frac{1}{2}\sum_{a\neq b}(\beta V)^2 q^2_{ab}\nonumber\\ &-& \ln
\left[ \int d\varepsilon_i P(\varepsilon_i ) \int Dc^{\dagger a}_i
Dc^a_i \exp\left\{ -S_{eff} (i) \right\}\right],
\end{eqnarray}
with $S_{eff} (i)$ given by Eq. (2). The self-consistency
conditions, Eqs. (4-6) then follow from

\begin{equation}
0=\delta F/\delta G (\tau ,\tau ' ); \; 0=\delta F/\delta \chi
(\tau ,\tau ' ); \; 0=\delta F/\delta q_{ab}.
\end{equation}

We stress that Eqs. (3-5) have been derived for the model with
{\em uniform} hopping elements $t_{ij}$ and interaction amplitudes
$V_{ij}$, in the $z\rightarrow\infty$ limit, but the {\em same}
equations hold for an infinite range model where these parameters
are random variables.

\subsection{The glass transition}

In our electronic model, the random site energies $\varepsilon_i$
play a role of static random fields. As a result, in presence of
disorder, the Edwards-Anderson  parameter $q_{EA}$ remains nonzero
for any temperature, and thus cannot serve as an order parameter.
To identify  the glass transition, we search for a replica
symmetry breaking (RSB) instability, following standard methods
\cite{re:AT,re:Mezard92}. We define $\delta q_{ab} = q_{ab} -q$,
and expand the free energy functional of Eq. (6) around the RS
solution. The resulting quadratic form (Hessian matrix) has the
matrix elements given by
\begin{eqnarray}
\frac{\partial^2 F}{\partial q_{ab} \partial q_{cd}}  = (\beta
V)^2 \delta_{ac}\delta_{bd} &-&V^4\int_{0}^{\beta}\int_{0}^{\beta}
\int_{0}^{\beta}\int_{0}^{\beta} d\tau_1 d\tau_2 d\tau_3 d\tau_4
[< \delta n _a (\tau_1 ) \delta n _b (\tau_2 ) \delta n _c (\tau_3
) \delta n _d (\tau_4 )>_{RS}\nonumber
\\&-& < \delta n _a (\tau_1 ) \delta n _b (\tau_2 )>_{RS} < \delta n_c
(\tau_3 )  \delta n _d (\tau_4 )>_{RS}],\label{eq:ch2_hessian}
\end{eqnarray}
where the expectation values are calculated in the RS solution.
Using standard manipulations\cite{re:AT}, and after lengthy
algebra, we finally arrive at the desired RSB stability criterion
that takes the form

\begin{equation}
1 = V^2 \left[\left(\chi_{loc} (\varepsilon_i
)\right]^2\right]_{dis}.\label{rsb}
\end{equation}
Here, $[...]_{dis}$ indicates the average over disorder, and
$\chi_{loc} (\varepsilon_i )$ is the {\em local compressibility},
that can be expressed as
\begin{equation}
\chi_{loc} (\varepsilon_i )= \frac{\partial}{\partial\varepsilon_i
} \frac{1}{\beta}\int_{o}^{\beta}d\tau <\delta n_i (\tau )>,
\end{equation}
and which is evaluated by carrying out quantum averages for a
fixed realization of disorder. The relevant expectation values
have to be carried with respect to the full local effective action
$S_{eff} (i)$ of Eq. (2), evaluated in the RS theory. In general,
the required computations cannot be carried out in close form,
primarily due to the unknown ``memory kernel'' $\chi(\tau -\tau
')$. However, as we will see, the algebra simplifies in several
limits, where explicit expressions can be obtained.

\section{Classical electron glass}

In the classical $(t=0)$ limit, the problem can easily be solved
in close form. We first focus on the replica symmetric (RS)
solution, and set $q_{ab}=q$ for all replica pairs. The
corresponding equation reads

\begin{equation}
q = \frac{1}{4}\int_{-\infty}^{+\infty}\frac{dx}{\sqrt{\pi}}
e^{-x^2/2}\tanh^2\left[ \frac{1}{2}x \left( (\beta V)^2 q + (\beta
W)^2\right)^{1/2}\right],
\end{equation}
where we have considered a Gaussian distribution of random site
energies of variance $W^2$. Note that the interactions introduce
an effective, {\em enhanced} disorder strength
\begin{equation}
W_{eff}=\sqrt{W^2 +V^2q}\label{eq:rescaled_disorder},
\end{equation}
since the frozen-in density fluctuations introduce an added
component to the random potential seen by the electron. As
expected, $q\neq 0$ for any temperature when $W\neq 0$. If the
interaction strength is appreciable as compared to disorder, we
thus expect the resistivity to display an appreciable {\em
increase} at low temperatures. We emphasize that this mechanism is
{\em different} from Anderson localization, which is going to be
discussed in the next section, but which also gives rise to a
resistivity increase at low temperatures.

Next, we examine the instability to glassy ordering. In the
classical ($t=0$) limit Eq. (\ref{rsb}) reduces to

\begin{equation}
1=\frac{1}{16}(\beta V)^2
\int_{-\infty}^{+\infty}\frac{dx}{\sqrt{\pi}}
e^{-x^2/2}\cosh^{-4}\left[ \frac{1}{2}x \beta W_{eff}(q)\right],
\end{equation}
with $W_{eff}(q)$ given by Eq.(\ref{eq:rescaled_disorder}).  The
resulting RSB instability line separates a low temperature glassy
phase from a high temperature ``bad metal'' phase. At large
disorder, these experssions simplify, and we find

\begin{equation}
T_G \approx \frac{1}{6\sqrt{2\pi}}\frac{V^2}{W},\;
W\rightarrow\infty .
\end{equation}
We conclude that $T_G$ decreases at large disorder. This is to be
expected, since in this limit the electrons drop in the lowest
potential minima of the random potential. This defines a unique
ground state, suppressing the {\em frustration} associated with
the glassy ordering, and thus reducing the glassy phase. It is
important to note that for the well known de Almeida-Thouless (AT)
line $T_{RSB}$ decreases {\em exponentially} in the strong field
limit. In contrast, we find that in our case, $T_G\sim 1/W$
decreases only slowly in the strong disorder limit. This is
important, since the glassy phase is expected to be most relevant
for disorder strengths sufficient to suppress uniform ordering. At
the same time, glassy behavior will only be observable if the
associated glass transition temperature remains appreciable.

\subsection{The glassy phase}

To understand this behavior, we investigate the structure of the
low-temperature glass phase. Consider the single-particle density
of states at T=0, which in the classical limit can be expressed as
\begin{equation}
\overline{\rho} (\varepsilon , t=0)= \frac{1}{N} \sum_i \delta
(\varepsilon -\varepsilon^{R}_i ),
\end{equation}
where $\varepsilon_i^{R} \equiv \varepsilon_i + \sum V_{ij} n_j$
are the renormalized site energies.  In the thermodynamic limit,
this quantity is nothing but the probability distribution $P_R
(\varepsilon^{R}_i )$.  It is analogous to the ``local field
distribution'' in the spin-glass models, and can be easily shown
to reduce to a simple Gaussian distribution in the RS theory,
establishing the {\em absence} of any gap for $T>T_G$. Obtaining
explicit results from a replica calculation in the glass phase is
more difficult, but useful insight can be achieved by using
standard simulation methods \cite{re:Palmer79,re:Pazmandi99} on
our equivalent infinite-range model; some typical results are
shown in Fig. 1.
\begin{figure}
\begin{center}
\includegraphics[width=4in]{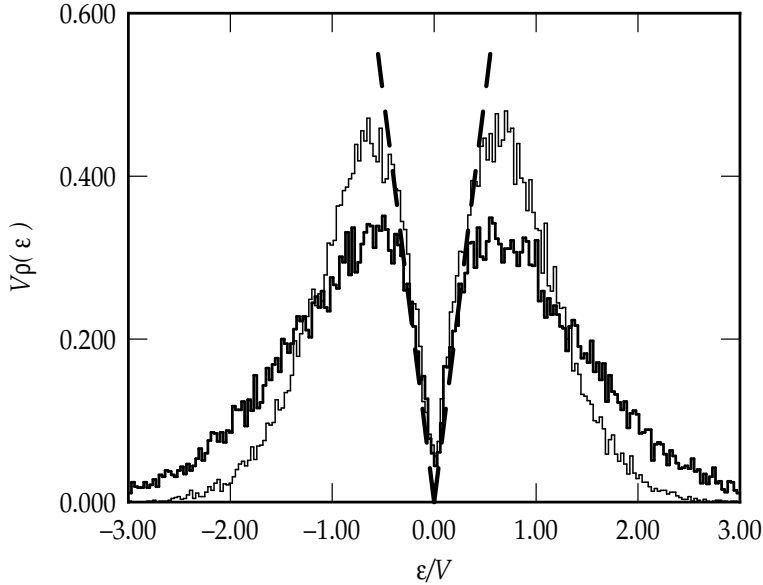}
\caption{Single particle density of states in the classical
($t=0$) limit at $T=0$, as a function of disorder strength.
Results are shown from a simulation on $N=200$ site system, for
$W/V = 0.5$ (thin line) and $W/V = 1.0$ (full  line). Note that
the low energy form of the gap takes a {\em universal} form,
independent of the disorder strength $W$. The dashed line follows
Eq. (16).}
\end{center}
\end{figure}
We find that as a result of glassy freezing, a pseudo-gap emerges
in the single-particle density of states, reminiscent of the
Coulomb gap of Efros and Shklovskii (ES) \cite{re:Efros75}. The
low energy form of this gap appears {\em universal},
\begin{equation}
\rho (\varepsilon)  \approx C\varepsilon^{\alpha}/V^2; \;\;\;
C=\alpha = 1,
\end{equation}
independent of the disorder strength $W$, again in striking
analogy with the predictions of ES. To establish this result, we
have used stability arguments very similar to those developed for
spin-glass (SG) models \cite{re:Pazmandi99}, demonstrating that
the form of Eq. (16) represents an exact {\em upper bound} for
$\rho(\varepsilon )$. For {\em infinite-ranged} SG models, as in
our case, this bound appears to be {\em saturated}, leading to
universal behavior. Such universality is often associated with a
critical, self-organized state of the system. Recent work
\cite{re:Pazmandi99} finds strong numerical evidence of such
criticality for SG models; we believe that the universal gap form
in our case has the same origin. Furthermore, assuming that the
universal form of Eq. (11) is obeyed immediately allows for an
estimate of $T_G (W)$.  Using Eq. (16) to estimate the gap size
for large disorder gives $T_G \sim E_g \sim V^2 /W$, in agreement
with Eq. (14).

The ergodicity breaking associated with the glassy freezing has
important consequences for our model. Again, using the close
similarity of our classical infinite range model to standard SG
models \cite{re:Mezard86}, it is not difficult to see that the
{\em zero-field cooled} (ZFC) compressibility vanishes at $T=0$,
in contrast to the field-cooled one, which remains finite.
Essentially, if the chemical potential is modified {\em after} the
system is cooled to $T=0$, the system immediately falls out of
equilibrium and displays hysteretic behavior \cite{re:Pazmandi99}
with vanishing {\em typical} compressibility. If this behavior
persists in finite dimensions and for more realistic Coulomb
interactions, it could explain the absence of screening in
disordered insulators.

\subsection{Arbitrary lattices and finite coordination: mean-field glassy
phase of the random-field Ising model.}

Simplest theories of glassy freezing \cite{re:Mezard86} are
obtained by examining models with random inter-site interactions.
In the case of disordered electronic systems, the interactions are
not random, but glassiness still emerges due to frustration
introduced by the competition of the interactions and disorder. As
we have seen for the Bethe lattice\cite{re:pd}, random
interactions are generated by renormalization effects, so that
standard DMFT approaches can still be used. However, one would
like to develop systematic approaches for arbitrary lattices and
in finite coordination. These issues already appear on the
classical level, where our model reduces to the random-field Ising
model (RFIM) \cite{re:Nattermann97}. To investigate the glassy
behavior of the RFIM, we developed\cite{re:Horbach01} a systematic
approach that can incorporate short-range fluctuation corrections
to the standard Bragg-Williams theory, following the method of
Plefka \cite{re:Plefka} and Georges et al. \cite{re:Georges90}.
This work has shown that:

\begin{itemize}
\item Corrections to even the lowest nontrivial order immediately
result in the appearance of a glassy phase for sufficiently strong
randomness.

\item This low-order treatment is sufficient in the joined limit
of large coordination and strong disorder.

\item The structure of the resulting glassy phase is characterized
by universal hysteresis and avalanche behavior emerging from the
self-organized criticality of the ordered state.
\end{itemize}

\section{Quantum melting of the electron glass}

Next, we investigate how the glass transition temperature can be
depressed by quantum fluctuations introduced by inter-site
electron tunneling. As in other quantum glass problems, quantum
fluctuations introduce dynamics in the problem, and the relevant
self-consistency equations cannot be solve in closed form for
general values of the parameters. In the following, we will see
that in the limit of large randomness, an exact solution is
possible.

\subsection{Quantum phase diagram}

The main source of difficulty in general quantum glass problems
relates to the existence of a self-consistently determined
``memory kernel'' $\chi (\tau -\tau ' )$ in the local effective
action. By the same reasoning as in the clasical case, one can
also ignore this term since this quantity is also bounded.

\begin{figure}[h]
\begin{center}
\includegraphics[width=4in]{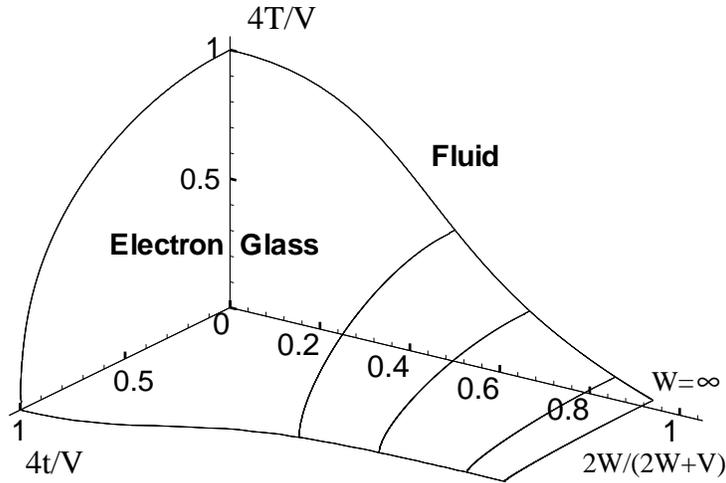}
\caption{Phase diagram as a function of quantum hopping $t$,
temperature $T$ and disorder strength $W$. Glass transition
temperature $T_G$ decreases only slowly (as $~1/W$) in the strong
disorder limit. In contrast, the critical value of the hoping
element $t_G$ remains finite as $W\rightarrow\infty$}
\end{center}
\end{figure}

The remaining action is that of \emph{noninteracting} electrons in
presence of a strong random potential. The resulting {\em  local}
compressibility then takes the form

\begin{equation}
\chi_{loc} (\varepsilon ) =\frac{\beta}{4}\int_{-\infty}^{+\infty}
d\omega \rho_{\varepsilon}(\omega ) \cosh^{-2}
(\frac{1}{2}\beta\omega ).
\end{equation}
Here,  $\rho_{\varepsilon}(\omega )$ is the local density of
states, which in the considered large $z$ limit is determined by
the solution of the CPA equation
\begin{equation}
\rho_{\varepsilon}(\omega )=-\frac{1}{\pi}{\rm Im} G(\omega);\;\;
G(\omega )= \int \frac{d\varepsilon P(\varepsilon )}{ \omega+i\eta
-\varepsilon -t^2 G(\omega )},
\end{equation}
In the limit $W/t >> 1$, it reduces to a narrow resonance of width
$\Delta =\pi t^2 P(0)\sim t^2/W$
\begin{equation}
\rho_{\varepsilon}(\omega ) \approx \frac{1}{\pi } \frac{\Delta
}{(\omega -\varepsilon )^2 +\Delta^2}.
\end{equation}

The resulting expression for the quantum critical line in the
large disorder limit takes the form
\begin{equation}
t_G (T=0, W\rightarrow\infty )= V/\sqrt{\pi }.
\end{equation}

At first glance, this result is surprising, since it means that a
{\em finite} value of the Fermi energy is required to melt the
electron glass at $T=0$, {\em even} in the $W\rightarrow\infty$
limit ! This is to be contrasted with the behavior of $T_G$ in the
classical limit, which according to Eq. (17) was found to decrease
as $1/W$ for strong disorder. At fist puzzling, the above result
in fact has a simple physical meaning. Namely, the small resonance
width (or ``hybridization energy'') $\Delta\sim t^2 /W$ can be
interpreted \cite{re:Anderson58,re:Dobrosavljevic97b} as the
characteristic energy scale for the electronic motion. As first
pointed by Anderson \cite{re:Anderson58}, according to Fermi's
golden rule, the transition rate to a neighboring site is
proportional to $\Delta$ and not $t$, and thus becomes extremely
small at large disorder. Thus the ``size'' of quantum
fluctuations, that replace the thermal fluctuations at $T=0$, is
proportional to $\delta\sim 1/W$, and thus becomes very small in
the large $W$ limit. We can now easily understand the qualitative
behavior shown in Eq. (24) by replacing $T\rightarrow \Delta\sim
t^2 /W$ in Eq. (17). The leading $W$ dependence {\em cancels out},
and we find a {\em finite} value for $t_G$ in the
$W\rightarrow\infty$ limit.

More generally, we can write an expression for the glass
transition critical line in the large disorder limit, as a
function of $\beta = 1/T$ and $t$ in the scaling form

\begin{equation}
1=(V/t)^2 \phi (\beta t^2 /W),
\end{equation}
with
\begin{equation}
\phi(z) = \frac{1}{4} z^2 \int_{-\infty}^{+\infty} dx \left[
\int_{-\infty}^{+\infty} dy \frac{1}{\pi} \frac{1}{1+ (x-y)^2}
\cosh^{-2} (\frac{1}{2} zy )\right]^2 .
\end{equation}

At finite disorder an exact solution is not possible, but we can
make analytical progress  motivated by our discussion of the large
$W$ limit. Namely, one can imagine evaluating the required local
compressibilities in Eq. (13) by a ``weak coupling'' expansion in
powers of the interaction $V$. To leading order, this means
evaluating the compressibilities at $V=0$, an approximation which
becomes exact for $W$ large. Such an approximation can be tested
for other spin glass problems. We have carried out the
corresponding computations for the infinite range Ising spin glass
model in a transverse field, where the exact critical transverse
field is known from numerical studies. We can expect the leading
approximation to {\em underestimate} the size of the glassy
region, i. e. the critical field, since the omitted ``memory
kernel'' introduces long range correlations in time, which make
the system more ``classical''. Indeed, we find that the leading
approximation underestimates the critical field by only about
30\%, whereas the next order correction makes an error of less
than 5\%. Encouraged by these arguments, we use this
``weak-coupling'' approximation for arbitrary disorder strength
$W$. Again, the computation of the compressibility reduces to that
of noninteracting electrons in a CPA formulation; the resulting
phase diagram is shown in Fig. 4.

\subsection{Quantum critical behavior of the electron glass}

So far, we have seen how our extended DMFT equations can be
simplified for large disorder, allowing an exact computation of
the phase boundary in this limit.  In our case, this quantum
critical line separates a (non-glassy) Fermi liquid phase, and a
metallic glass phase which, as we will see, features non-Fermi
liquid behavior. If one is interested in details of {\em dynamics}
of the electrons near the quantum critical line, the above
simplifications do not apply, and one is forced to
self-consistently calculate the form of the "memory kernel" (local
dynamic compressibility) $\chi (\tau -\tau ')$. Fortunately, this
task can be carried out using methods very similar to those
developed for DMFT models for metallic spin
glasses\cite{re:Read95}. Formulating such a theory is technically
possible because the exact quantum critical behavior is captured
when the relevant field theory is examined at the Gaussian
level\cite{re:Miller93}, in the considered limit of large
dimensions.

Because of technical complexity of this calculation, we only
report the main results, while the details can be found in Ref.
[\citenum{Denis}]. In this paper, the full replica-symmetry broken
(RSB) solution was found both around the quantum critical line and
in the glassy phase. In the Fermi liquid phase, the memory kernel
was fond to take the form
\[
V^2 \chi (\omega_n )=D(\omega_n)+\beta q_{{\rm
EA}}\delta_{\omega_n,0}, \]
with
 \[
D(\omega_n)= -y q_{{\rm EA}}^2 /V^4 -\sqrt{|\omega_n|+\Delta}.\]
Here, $\Delta$ is a characteristic energy scale that vanishes on
the critical line, which also determines a crossover temperature
scale separating the Fermi liquid from the quantum-critical
regime. In contrast to conventional quantum critical phenomena,
but similarly as in metallic spin glasses, the "gap" scale $\Delta
=0$ not only on the critical line, but remains zero
\emph{throughout the entire glassy phase}. As a result, the
excitations in this region assume a non-Fermi liquid form
\[ D(\omega_n)= -y q_{{\rm EA}}^2 /V^4 -\sqrt{|\omega_n|}.\]
This behavior reflects the emergence of soft "replicon"
modes\cite{re:Mezard86} describing in our case represent low
energy charge rearrangements inside the glassy phase. At finite
temperatures, electrons undergo inelastic scattering from such
collective excitations, leading to the temperature dependence of
the resistivity that takes the following non-Fermi liquid form
\[\rho (T) =\rho(o) +AT^{3/2}.\]
Interestingly, very recent experiments\cite{SBPRL02} on two
dimensional electron gases in silicon have revealed precisely such
temperature dependence of the resistivity. This behavior has been
observed in what appears to be an intermediate metallic glass
phase separating a conventional (Fermi liquid) metal at high
carrier density, from an insulator at the lowest densities.

Another interesting feature of the predicted quantum critical
behavior relates to disorder dependence of the crossover exponent
$\phi$ describing how the gap scale $\Delta\sim\delta r^{\phi}$
vanishes as a function of the distance $\delta r$ from the
critical line. Calculations\cite{arrachea}  show that $\phi =2$ in
presence of site energy disorder, which for our model plays a role
of a random symmetry breaking field, and $\phi =1$ in its absence.
This indicates that site disorder, which is common in disordered
electronic systems, produces a particularly large quantum critical
region, which could be the origin of large dephasing observed in
many materials near the metal-insulator transition.

\subsection{Effects of Anderson localization}

As we have seen, the stability of the glassy phase is crucially
determined by the electronic mobility at $T=0$.  More precisely,
we have shown that the relevant energy scale that determines the
size of quantum fluctuations introduced by the electrons is given
by the local ``resonance width'' $\Delta$. It is important to
recall that precisely this quantity may be considered
\cite{re:Anderson58} as an order parameter for Anderson
localization of noninteracting electrons. Very recent work
\cite{re:Dobrosavljevic97b,re:Dobrosavljevic98} demonstrated that
the {\em typical} value of this quantity plays the same role even
at a Mott-Anderson transition. We thus expect $\Delta$ to
generally vanish in the insulating state. As a result, we expect
the stability of the glassy phase to be strongly affected by
Anderson localization effects, as we will explicitly demonstrate
in the next section.

\section{Glassy behavior near the Mott-Anderson transition}

On physical grounds, one expects the quantum fluctuations
\cite{pastor} associated with mobile electrons \ to suppress
glassy ordering, but their precise effects remain to be
elucidated. Note that even the \emph{amplitude }of such quantum
fluctuations must be a singular function of the distance to the
MIT, since they are dynamically determined by processes that
control the electronic mobility.

To clarify the situation, the following basic questions need to be
addressed: (1) Does the MIT\ coincide with the onset of glassy
behavior? (2) How do different physical processes that can
localize electrons \ affect the stability of the glass phase? In
the following, we provide simple and physically transparent \
answers to both questions. We find that: (a) Glassy behavior
generally emerges before the electrons localize; (b) Anderson
localization \cite{re:Anderson58} enhances the stability of the
glassy phase, while Mott localization \cite{re:Mott90} tends to
suppress it.

In order to be able to examine both the effects of Anderson and
Mott localization, we concentrate on extended Hubbard models given
by the Hamiltonian

\[
H=\sum_{ij\sigma}(-t_{ij}+\varepsilon_{i}\delta_{ij})c_{i,\sigma}^{\dagger
}c_{j,\sigma}+U\sum_{i}n_{i\uparrow}n_{i\downarrow}+\sum_{ij}V_{ij}\delta
n_{i}\delta n_{j}.
\]
Here, $\delta n_{i}=$ $n_{i}-$ $\left\langle n_{i}\right\rangle $
represent local density fluctuations ( $\left\langle
n_{i}\right\rangle $ is the site-averaged electron density), $U$
is the on-site interaction, and $\varepsilon_{i\ }$are$_{\
}$Gaussian distributed random site energies of variance $W^{2}$.
In order to allow for glassy freezing of electrons in the charge
sector, we introduce weak inter-site density-density interactions
$V_{ij}$, which we also also choose to be Gaussian distributed
random variables of variance $V^{2}$ /$z$ ($z$ is the coordination
number). We emphasize that, in contrast to previous work
\cite{re:pd}, we shall now keep the coordination number $z$
finite, in order to allow for the possibility of Anderson
localization. To investigate the emergence of glassy ordering, we
formally average over disorder by using standard replica methods
\cite{Darko}, and introduce collective $Q$-fields to decouple the
inter-site $V$-term \cite{Darko}. A mean-field is then obtained by
evaluating the $Q$-fields at the saddle-point level. The resulting
stability criterion takes the form similar as before

\begin{equation}
1-V^{2}%
{\displaystyle\sum\limits_{j}}
[\chi_{ij}^{2}]_{dis}=0. \label{instab}%
\end{equation}
Here, the non-local static compressibilities are defined (for a
fixed realization of disorder) as
\begin{equation}
\chi_{ij} =-\partial n_{i}/\partial\varepsilon_{j},%
\end{equation}
where $n_{i}$ is the local quantum expectation value of the
electron density, and $[\cdots]_{dis}$ represents the average over
disorder. Obviously, the stability of the glass phase is
determined by the behavior of the four-order correlation function
$\chi^{(2)} =\sum\limits_{j} [\chi_{ij}^{2}]_{dis}$ in the
vicinity of the metal-insulator transition. We emphasize that this
quantity is to be calculated in a disordered Hubbard model with
finite range hopping, i.e. in the vicinity of the Mott-Anderson
transition. The critical behavior of $\chi^{(2)}$ is very
difficult to calculate in generaql, but we will see that simple
results can be obtained in the limits of weak and strong disorder,
as follows.

\subsection{Large disorder}

As the disorder grows, the system approaches the Anderson
transition at $t=t_{c}(W)\sim W$. The first hint of singular
behavior of $\chi^{(2)}$ in an Anderson insulator is seen by
examining the deeply insulating, i. e. atomic limit $W\gg t,$
where to leading order we set $t=0$ and obtain
$\chi_{ij}=\delta(\varepsilon_{i}-\mu )\delta_{ij}$, i.e.
$\chi^{(2)}=[\delta^{2}(\varepsilon_{i}-\mu)]_{dis}=+\infty$
diverges! Since we expect all quantities to behave in
qualitatively the same fashion throughout the insulating phase, we
anticipate $\chi^{(2)}$ to diverge already at the Anderson
transition. Note that, since the instability of the glassy phase
occurs already at $\chi^{(2)}=V^{-2}$, the glass transition must
\emph{precede} the localization transition. Thus, for any finite
inter-site interaction $V$, we predict the emergence of an
intermediate \emph{metallic glass phase }separating the Fermi
liquid from the Anderson insulator. Assuming
that near the transition%

\begin{equation}
\chi^{(2)}\simeq\frac{A}{W^{2}}((t/W)-B)^{-\alpha}%
\end{equation}
($A$ and $B=t_{c}/W$ are constants of order unity), from Eq.
$\left( \ref{instab}\right)  $ we can estimate the form of the
glass transition line, and we get
\begin{equation}
\delta t(W)=t_{G}(W)-t_{c}(W)\sim
V^{2/\alpha}W^{1-2/\alpha};\;W\rightarrow
\infty. \label{anderson}%
\end{equation}
The glass transition and the Anderson transition lines are
predicted to converge at large disorder for $\alpha<2,$ and
diverge for $\alpha>2$. Since all the known exponents
characterizing the localization transition seem to grow with
dimensionality, we may expect a particularly large metallic glass
phase in large dimensions.

\subsubsection{Anderson localization on Bethe lattice}

\textit{ }In order to confirm this scenario by explicit
calculations, we compute the behavior of $\chi^{(2)}$ at the
Anderson transition of a half-filled Bethe lattice of coordination
$z=3.$We use an essentially exact numerical approach
\cite{re:Dobrosavljevic97b} based on the recursive structure of
the Bethe lattice \cite{re:Abou-Chacra73}. In this approach, local
and non-local Green's functions on a Bethe lattice can be sampled
from a large ensemble, and the compressibilities $\chi_{ij}$ can
be then calculated by examining how a local charge density $n_{i}$
is modified by an infinitesimal variation of the local site energy
$\varepsilon_{j}$ on another site. To do this, we have taken
special care in evaluating the local charge densities $n_{i}$ by
numerically computing the required frequency summations over the
Matsubara axis, where the numerical difficulties are minimized.
Using this method, we have calculated $\chi^{(2)}$as a function of
$W/t$(for this lattice at half-filling $E_{F}=$ 2$\sqrt{2}t$~),
and find that it decreases exponentially \cite{re:Mirlin00} as the
Anderson transition is approached. We emphasize that only a finite
enhancement of $\chi^{(2)}$ is required to trigger the instability
to glassy ordering, which therefore occurs well before the
Anderson transition is reached. The resulting $T=0$ phase diagram,
valid in the limit of large disorder, is presented in Fig. 1. Note
that the glass transition line in this case has the form
$t_{G}(W)\sim W$, in agreement with the fact that exponential
critical behavior of $\chi^{(2)}$ corresponds to $\alpha
\rightarrow\infty$ in the above general scenario. These results
are strikingly different from those obtained in a theory which
ignores localization \cite{re:pd}, where $t_{G}(W)$ was found to
be weakly dependent on disorder, and remain\emph{ finite} as
$W\longrightarrow\infty$. Anderson localization effects thus
strongly enhance the stability of the glass phase at sufficiently
large disorder. Nevertheless, since the Fermi liquid to metallic
glass (FMG) transition occurs at a finite distance \emph{before}
the localization transition, we do not expect the leading quantum
critical behavior \cite{Denis} at the FMG transition to be
qualitatively modified by the localization effects.

\begin{figure}[ptb]
\begin{center}
\includegraphics[width=4in ]{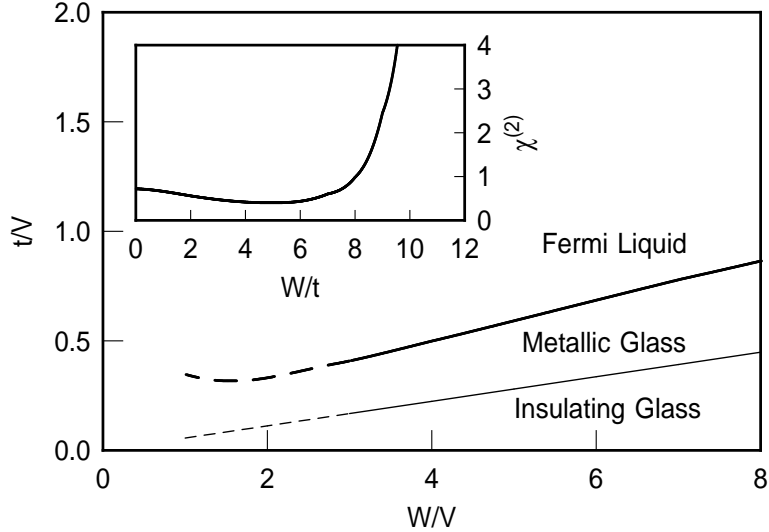}
\end{center}
\caption{Phase diagram for the $z=3$ Bethe lattice, valid in the
large
disorder limit. The inset shows $\chi^{(2)}$ as a function of disorder $W$.}%
\label{fig1}%
\end{figure}

\subsubsection{Typical medium treatment of Anderson localization}
\label{sec:ch6_tmm}

As an alternative approach to the Bethe lattice calculation, in
this section we introduce Anderson localization to the problem by
using the formalism of "Typical Medium Theory"\cite{tmt} (TMT). We
calculate the cavity field $\Delta_{TYP}(\o)$ by solving the
relevant self-consistency condintion\cite{tmt}, which in turn
allows us to find local compressibilities: \bea
 \chi_{ii}&=&-{\dd n\over\dd \ve_i}={1\over\pi}{\dd\over\dd\ve_i}\int_{-\infty}^0
 d\o Im G(\ve_i,\o,W)\\
  G(\ve_i,\o,W)&=&{1\over\o-\ve_i-\Delta_{TYP}(\o)},
\eea needed to determine the critical line of the glass
transition. These calculations were performed using a model of
semicircular bare DOS $\rho_0(\o)$  and box distribution of
disorder $P(\ve_i)$. The resulting phase diagram is shown in Fig.
6.
\begin{figure}
\begin{center}
\includegraphics[width=4in]{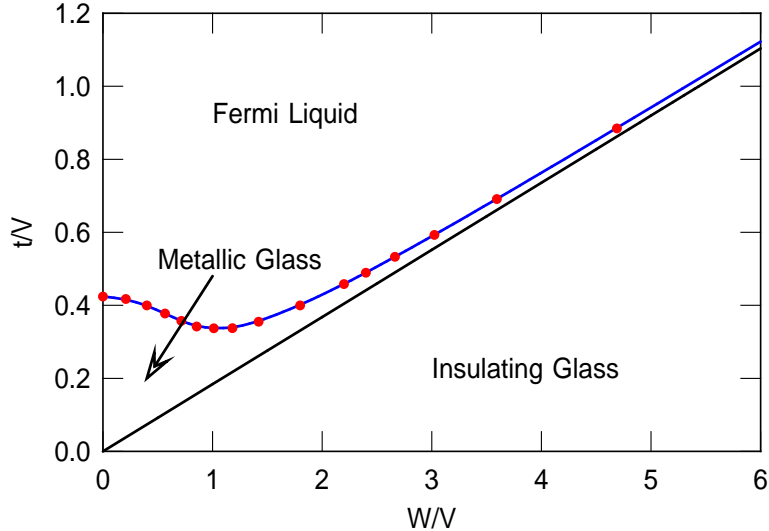}\caption{Phase
diagram from Typical Medium Theory of Anderson
localization\cite{tmt}, giving $\a =1$. The intermediate metallic
glassy phase shrinks as disorder $W$ grows, as expected. Compare
this to the Bethe lattice case Fig. 5, where $\a=\infty$.
\label{arbitrary_anderons_phase}}
\end{center}
\end{figure}
The intermediate  metallic glassy phase still exists, but shrinks
as $W\rightarrow\infty$, reflecting the small value of the
critical exponent $\alpha=1$, which can be shown analytically
within TMT. A more realistic vales of this exponent, corresponding
to $d=3$ require more detailed numerical calculations, which
remains a challenge for future work.

\subsection{Low disorder - Mott transition}

In the limit of weak disorder $W\ll U,V$, and interactions drive
the metal-insulator transition. Concentrating on the model at
half-filling, the system will undergo a Mott transition
\cite{re:Mott90} as the hopping $t$ is sufficiently reduced. Since
for the Mott transition $t_{Mott}(U)\sim U$, near the transition
$W\ll t$, and to leading order we can ignore the localization
effects. In addition, we assume that $V\ll U,$ and to leading
order the compressibilities have to be calculated with respect to
the action $S_{el}$ of a disordered Hubbard model. The simplest
formulation that can describe the effects of weak disorder on such
a Mott transition is obtained from the dynamical mean-field theory
(DMFT) \cite{re:Georges96}. This formulation, which ignores
localization effects, is obtained by rescaling the hopping
elements $t\rightarrow t/\sqrt{z}$ and then formally taking the
limit of large coordination $z\rightarrow\infty$. To obtain
qualitatively correct analytical results describing the vicinity
of the disordered Mott transition at $T=0,$ we have solved the
DMFT equations using a 4-boson method\cite{Darko}. At weak
disorder, these equations can be easily solved in close form, and
we simply report the relevant results. The critical value of
hopping for the Mott transition is
found to decrease with disorder, as%
\begin{equation}
t_{c}(W)\approx t_{c}^{o}\,(1-4(W/U)^{2}+\cdots),
\end{equation}
where for a simple semi-circular density of states
\cite{re:Georges96} $t_{c}^{o}=3\pi U/64$ (in this model, the
bandwidth $B=4t$). Physically, the disorder tends to suppress the
Mott insulating state, since it broadens the Hubbard bands and
narrows the Mott-Hubbard gap. At sufficiently strong disorder
$W\geq U$, the Mott insulator is suppressed even in the atomic
limit $t\rightarrow0$. The behavior of the compressibilities can
also be calculated near the Mott
transition, and to leading order we find%
\begin{equation}
\chi^{(2)}=\left[  \frac{8}{3\pi
t_{c}^{o}}(1-\frac{t_{c}(W)}{t})\right] ^{2}(1+28(W/U)^{2}).
\end{equation}
Therefore, as any compressibility, $\chi^{(2)}$ is found to be
very small in the vicinity of the Mott transition, even in
presence of finite disorder. As a result, the tendency to glassy
ordering is strongly suppressed at weak disorder, where one
approaches the Mott insulating state.
\begin{figure}
[ptbh]
\begin{center}
\includegraphics[width=4in]
{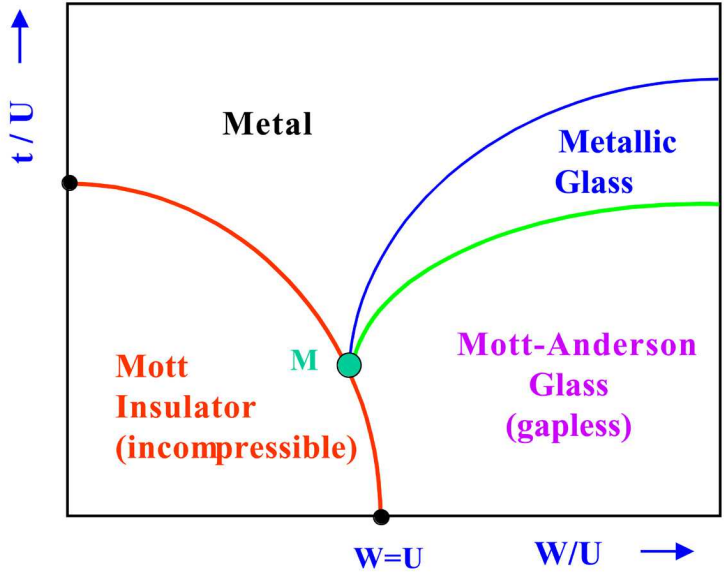}%
\caption{Schematic phase diagram for an extended Hubbard model
with disorder, as a function of the hopping element $t$ and the
disordered strength $W$, both expressed in units of the on-site
interaction $U$. The size of the metallic glass phase is
determined by the strength of the inter-site interaction $V$.}%
\end{center}
\end{figure}

Finally, having analyzed the limits of weak and strong disorder,
we briefly comment on what may be expected in the intermediate
region $W\sim U$. On general grounds, we expect a global phase
diagram as shown in Fig.7. The Mott gap cannot exist for $W>U$, so
in this region and for sufficiently small $t$ (i. e. kinetic
energy), one enters an gapless (compressible) Mott-Anderson
insulator. For $W\sim U,$ the computation of $\chi^{(2)}$ requires
the full solution of the Mott-Anderson problem. The required
calculations can and should be performed using the formulation of
Ref. \citenum{re:Dobrosavljevic97b,re:Dobrosavljevic98}, but that
difficult task is a challeng for the future. However, based on
general arguments presented above, we expect $\chi^{(2)}$ to
\emph{vanish} as one approaches the Mott insulator $(W<U)$, but to
\emph{diverge} as one approaches the Mott-Anderson insulator
($W>U).$ Near the tetracritical point M (see Fig. 2), we may
expect $\chi^{(2)}\sim\delta W^{-\alpha}\delta t^{\beta },$ where
$\delta W=W-W_{Mott}(t)$ is the distance to the Mott transition
line, and $\delta t=t-t_{c}(W)$ is the distance to the
Mott-Anderson line. Using this ansatz and Eq. (\ref{instab}), we
find the glass transition line to
take the form%

\begin{equation}
\delta t=t_{G}(W)-t_{c}(W)\sim\delta W^{\beta/\alpha};\;W\sim
W_{M}.
\label{tetracritical}%
\end{equation}

We thus expect the intermediate metallic glass phase to be
suppressed as the disorder is reduced, and one approaches the Mott
insulating state. Physically, glassy behavior of electrons
corresponds to many low-lying rearrangements of the charge
density; such rearrangements are energetically unfavorable close
to the (incompressible) Mott insulator, since the on-site
repulsion $U$ opposes charge fluctuations. Interestingly, very
recent experiments on low density electrons in silicon MOSFETs
have revealed the existence of exactly such an intermediate
metallic glass phase in low mobility (highly disordered) samples
\cite{SBPRL02}. In contrast, in high mobility (low disorder)
samples \cite{JJPRL02}, no intermediate metallic glass phase is
seen, and glassy behavior emerges only as one enters the
insulator, consistent with our theory. Similar conclusions have
also been reported in studies of highly disordered InO$_{2}$ films
\cite{films21,films22,films23,films24,films25}, where the glassy
slowing down of the electron dynamics seems to be suppressed as
the disorder is reduced and one crosses over from an Anderson-like
to a Mott-like insulator. In addition, these experiments
\cite{SBPRL02,JJPRL02} provide striking evidence of
scale-invariant dynamical correlations inside the glass phase,
consistent with the hierarchical picture of glassy dynamics, as
generally emerging from mean-field approaches \cite{re:Mezard86}
such as the one used in this work. \pagebreak

\section{Conclusions}

Recent years have witnessed enormous renewed interest in the
metal-insulator transition. Scores of new and fascinating
materials are being fabricated, with properties that could not be
anticipated. A common theme in many of these systems is the
presence of both the strong electron-electron interactions and
disorder, a situation which proved difficult to analyze using
conventional theoretical methods. In this paper, we have described
a novel approach to this difficult problem, and shown that it can
capture most relevant processes. This formulation can easily be
adapted to many realistic situations and will open new avenues for
the development of materials science research.

\acknowledgments     

The author would like to acknowledge his collaborators A. A.
Pastor, M.H. Horbach, D. Tanaskovi\'c, D. Dalidovich, L. Arrachea,
and M.J. Rozenberg, with whom it was a pleasure to explore the
physics of electron glasses. I also thank S. Bogdanovich, S.
Chakravarty, J. Jaroszynski, D. Popovi\'{c}, Z. Ovadyahu, J.
Schmalian, and G. Zimanyi  for useful discussions. This work was
supported by the NSF grant DMR-9974311 and DMR-0234215, and the
National High Magnetic Field Laboratory.



\end{document}

%% file: mydefs.tex
\def\be{\begin{equation}}
\def\ee{\end{equation}}
\def\bea{\begin{eqnarray}}
\def\eea{\end{eqnarray}}
\def\bi{\begin{itemize}}
\def\ei{\end{itemize}}
\def\ve{\varepsilon}

\def\a{\alpha}

\def\o{\omega}

\def\1n{{\partial\over\partial n}}

\def\dd{\partial}

%% file: canari.bbl
\begin{thebibliography}{10}

\bibitem{re:Mott90}
N.~F. Mott, {\em Metal-Insulator transition}, Taylor \& Francis,
London, 1990.

\bibitem{re:Anderson58}
P.~W. Anderson, ``Absence of diffusion in certain random
lattices,'' {\em Phys.
  Rev.} {\bf 109}, pp.~1492--1505, 1958.

\bibitem{re:Sondhi97}
S.~Sondhi, S.~Girvin, J.~Carini, and D.~Shahar, ``Continuous
quantum phase
  transitions,'' {\em Rev. Mod. Phys.} {\bf 69}, p.~315, 1997.

\bibitem{re:Lee85}
P.~A. Lee and T.~V. Ramakrishnan, ``Disordered electronic
systems,'' {\em Rev.
  Mod. Phys.} {\bf 57}, p.~287, 1985.

\bibitem{eglass1}
J.~H. Davies, P.~A. Lee, and T.~M. Rice, ``Electron glass,'' {\em
Phys. Rev.
  Lett.} {\bf 49}, pp.~758--761, 1982.

\bibitem{eglass2}
M.~Pollak and A.~Hunt, ``Slow processes in disordered solids,'' in
{\em Hopping
  Transport in Solids},  M.~Pollak and B.~I. Shklovskii, eds., pp.~175--206,
  Elsevier, Amsterdam, 1991.

\bibitem{films21}
M.~Ben-Chorin, D.~Kowal, and Z.~Ovadyahu, ``Anomalous field effect
in gated
  {A}nderson insulators,'' {\em Phys. Rev. B} {\bf 44}, pp.~3420--3423, 1991.

\bibitem{films22}
M.~Ben-Chorin, Z.~Ovadyahu, and M.~Pollak, ``Nonequilibrium
transport and slow
  relaxation in hopping conductivity,'' {\em Phys. Rev. B} {\bf 48},
  pp.~15025--15034, 1993.

\bibitem{films23}
Z.~Ovadyahu and M.~Pollak, ``Disorder and magnetic field
dependence of slow
  electronic relaxation,'' {\em Phys. Rev. Lett.} {\bf 79}, pp.~459--462, 1997.

\bibitem{films24}
A.~Vaknin, Z.~Ovadyahu, and M.~Pollak, ``Evidence for interactions
in
  nonergodic electronic transport,'' {\em Phys. Rev. Lett.} {\bf 81},
  pp.~669--672, 1998.

\bibitem{films25}
A.~Vaknin, Z.~Ovadyahu, and M.~Pollak, ``Aging effects in an
{A}nderson
  insulator,'' {\em Phys. Rev. Lett.} {\bf 84}, pp.~3402--3405, 2000.

\bibitem{re:Georges96}
A.~Georges, G.~Kotliar, W.~Krauth, and M.~J. Rozenberg,
``Dynamical mean-field
  theory of strongly correlated fermion systems and the limit of infinite
  dimensions,'' {\em Rev. Mod. Phys.} {\bf 68}, p.~13, 1996.

\bibitem{re:Dobrosavljevic98}
V.~Dobrosavljevi\'{c} and G.~Kotliar, ``Dynamical mean-field
studies of
  metal-insulator transitions,'' {\em Phil. Trans. R. Soc. Lond. A} {\bf 356},
  p.~1, 1998.

\bibitem{re:Rozenberg96}
M.~J. Rozenberg, G.~Kotliar, and H.~Kajueter, ``Transfer of
spectral weight in
  spectroscopies of correlated electron systems,'' {\em Phys. Rev. B} {\bf 54},
  p.~8542, 1996.

\bibitem{re:Massey96}
J.~G. Massey and M.~Lee, ``Low-frequency noise probe of
interacting charge
  dynamics in variable-range hopping boron-doped silicon,'' {\em Phys. Rev.
  Lett.} {\bf 77}, p.~3399, 1996.

\bibitem{re:Efros75}
A.~L. Efros and B.~I. Shklovskii, ``Coulomb gap and low
temperature
  conductivity of disordered systems,'' {\em J. Phys. C} {\bf 8}, pp.~L49--51,
  1975.

\bibitem{films32}
G.~Martinez-Arizala, C.~Christiansen, D.~E. Grupp, N.~Markovic,
A.~M. Mack, and
  A.~M. Goldman, ``Coulomb-glass-like behavior of ultrathin films of metals,''
  {\em Phys. Rev. B} {\bf 57}, pp.~R670--R672, 1998.

\bibitem{re:Belitz95}
D.~Belitz and T.~R. Kirkpatrick, ``Anderson-{M}ott transition as a
  quantum-glass problem,'' {\em Phys. Rev. B} {\bf 52}, p.~13922, 1995.

\bibitem{newgang}
V.~Dobrosavljevi\'c, E.~Abrahams, E.~Miranda, and S.~Chakravarty,
``Scaling
  theory of two-dimensional metal-insulator transitions,'' {\em Phys. Rev.
  Lett.} {\bf 79}, pp.~455--458, 1997.

\bibitem{Sudip}
S.~Chakravarty, S.~Kivelson, C.~Nayak, and K.~Voelker, ``Wigner
glass,
  spin-liquids, and the metal-insulator transition,'' {\em Phil. Mag. B} {\bf
  79}, p.~859, 1999.

\bibitem{re:pd}
A.~A. Pastor and V.~Dobrosavljevi\'{c}, ``Melting of the electron
glass,'' {\em
  Phys. Rev. Lett.} {\bf 83}, p.~4642, 1999.

\bibitem{re:Mezard86}
M.~Mezard, G.~Parisi, and M.~A. Virasoro, {\em Spin Glass theory
and beyond},
  World Scientific, Singapore, 1986.

\bibitem{re:Dobrosavljevic94}
V.~Dobrosavljevi\'c and G.~Kotliar, ``Strong correlations and
disorder in
  d=$\infty$ and beyond,'' {\em Phys. Rev. B} {\bf 50}, p.~1430, 1994.

\bibitem{re:Dobrosavljevic97b}
V.~Dobrosavljevi\'c and G.~Kotliar, ``Mean field theory of the
  {M}ott-{A}nderson transition,'' {\em Phys. Rev. Lett.} {\bf 78}, p.~3943,
  1997.

\bibitem{re:Cummins94}
H.~Z. Cummins, G.~Li, W.~M. Du, and J.~Hernandez, ``Relaxational
dynamics in
  supercooled liquids: experimental tests of the mode coupling theory,'' {\em
  Physica A} {\bf 204}, p.~169, 1994.

\bibitem{re:AT}
J.~de~Almeida and D.J.Thouless, ``Stability of the
sherrington-kirkpatrick
  solution of a spin glass model,'' {\em J. Phys. A} {\bf 11}, p.~983, 1978.

\bibitem{re:Mezard92}
M.~Mezard and A.~P. Young, ``Replica symmetry breaking in the
random field
  ising model,'' {\em Europhys. Lett.} {\bf 18}, pp.~653--659, 1992.

\bibitem{re:Palmer79}
R.~G. Palmer and C.~M. Pond, ``Internal field distribution in
model spin
  glasses,'' {\em J. Phys. F} {\bf 9}, p.~1451, 1979.

\bibitem{re:Pazmandi99}
F.~Pazmandi, G.~Zarand, and G.~T. Zimanyi, ``Self-organized
criticality in the
  hysteresis of the {S}herrington-{K}irkpatrick model,'' {\em Phys. Rev. Lett.}
  {\bf 83}, pp.~1034--1037, 1999.

\bibitem{re:Nattermann97}
T.~Nattermann, ``Theory of the random field {I}sing model,'' {\em
  cond-mat/9705295} , 1997.

\bibitem{re:Horbach01}
A.~A. Pastor, V.~Dobrosavreljevi\'c, and M.~L. Horbach,
``Mean-field glassy
  phase of the random field {I}sing model,'' {\em Phys. Rev. B} {\bf 66},
  p.~014413(14), 2001.

\bibitem{re:Plefka}
T.~Plefka, ``Convergence condition of the tap equation for the
infinite-ranged
  {I}sing spin glass model,'' {\em J. Phys. A} {\bf 15}, pp.~1971--1978, 1982.

\bibitem{re:Georges90}
A.~Georges, M.~Mezard, and J.~S. Yedida, ``Low-temperature phase
of the {I}sing
  spin glass on a hypercubic lattice,'' {\em Phys. Rev. Lett.} {\bf 64},
  p.~2937, 1990.

\bibitem{re:Read95}
N.~Read, S.~Sachdev, and J.~Ye, ``Landau theory of quantum spin
glasses of
  rotors and {I}sing spins,'' {\em Phys. Rev. B} {\bf 52}, p.~384, 1995.

\bibitem{re:Miller93}
J.~Miller and D.~A. Huse, ``Zero-temperature critical behavior of
the
  infinite-range quantum {I}sing spin glass,'' {\em Phys. Rev. Lett.} {\bf 70},
  p.~3147, 1993.

\bibitem{Denis}
D.~Dalidovich and V.~Dobrosavljevi\'c, ``Landau theory of the
{F}ermi-liquid to
  electron-glass transition,'' {\em Phys. Rev. B} {\bf 66}, p.~081107(4), 2002.

\bibitem{SBPRL02}
S.~Bogdanovich and D.~Popovi\'c, ``Onset of glassy dynamics in a
  two-dimensional electron system in silicon,'' {\em Phys. Rev. Lett.} {\bf
  88}, p.~236401(4), 2002.

\bibitem{arrachea}
L.~Arrachea, D.~Dalidovich, V.~Dobrosavljevi\'c, and M.~J.
Rozenberg, ``Melting
  transition of an {I}sing glass driven by magnetic field,'' {\em Phys. Rev. B
  (in press)} , 2004.

\bibitem{pastor}
A.~A. Pastor and V.~Dobrosavljevi\'c, ``Melting of the electron
glass,'' {\em
  Phys. Rev. Lett.} {\bf 83}, pp.~4642--4645, 1999.

\bibitem{Darko}
V.~Dobrosavljevi\'c, D.~Tanaskovi\'c, and A.~A. Pastor, ``Glassy
behavior of
  electrons near metal-insulator transitions,'' {\em Phys. Rev. Lett.} {\bf
  90}, p.~016402(4), 2003.

\bibitem{re:Abou-Chacra73}
R.~Abou-Chacra, P.~W. Anderson, and D.~Thouless, ``A
selfconsistent theory of
  localization,'' {\em J. Phys. C} {\bf 6}, p.~1734, 1973.

\bibitem{re:Mirlin00}
A.~D. Mirlin, ``Statistics of energy levels and eigenfunctions in
disordered
  systems,'' {\em Phys. Rep.} {\bf 326}, p.~259, 2000.

\bibitem{tmt}
V.~Dobrosavljevi\'c, A.~Pastor, and B.~K. Nikoli\'c, ``Typical
medium theory of
  {A}nderson localization: A local order parameter approach to strong disorder
  effects,'' {\em Europhys. Lett.} {\bf 62}, pp.~76--82, 2003.

\bibitem{JJPRL02}
J.~Jaroszy\'nski, D.~Popovi\'c, and T.~M. Klapwijk, ``Universal
behavior of the
  resistance noise across the metal-insulator transition in silicon inversion
  layers,'' {\em Phys. Rev. Lett.} {\bf 89}, p.~276401(4), 2002.

\end{thebibliography}
